\documentclass[a4paper, amsfonts, amssymb, amsmath, reprint, showkeys, nofootinbib, twoside]{revtex4-1}
\usepackage[english]{babel}
\usepackage[utf8]{inputenc}
\usepackage[colorinlistoftodos, color=green!40, prependcaption]{todonotes}
\usepackage{amsthm}
\usepackage{xcolor}
\usepackage{graphicx}
\usepackage[T1]{fontenc}
\usepackage[pdftex, pdftitle={Article}, pdfauthor={Author}]{hyperref} 
\begin{document}
\title{Generalized second law of thermodynamics for the matter creation scenario and emergence of phantom regime}

\author{V\'ictor H. C\'ardenas$^1$}
\email{victor.cardenas@uv.cl}

\author{Miguel Cruz$^2$}
\email{miguelcruz02@uv.mx}

\author{Samuel Lepe$^3$}
\email{samuel.lepe@pucv.cl}

\affiliation{$^1$Instituto de F\'{\i}sica y Astronom\'ia, Universidad de Valpara\'iso, Gran Breta\~na 1111, Valpara\'iso, Chile \\
$^2$Facultad de F\'{\i}sica, Universidad Veracruzana 91097, Xalapa, Veracruz, M\'exico \\
$^3$Instituto de F\'{\i}sica, Facultad de Ciencias, Pontificia Universidad Cat\'olica de Valpara\'\i so, Avenida Brasil 2950, Valpara\'iso, Chile 
}

\date{\today} 

\begin{abstract}
This work is focused on the exploration of the thermodynamics foundations of the matter creation scenario when a generalized form of the second law of thermodynamics for this scheme is implemented. In this scenario we consider an expanding cosmology in which the produced particles are trapped by the apparent horizon. The scheme leads to phantom evolution but at first glance it lacks of physical consistency. However, we show explicitly that the inclusion of chemical potential into the description solves the thermodynamics issues of the model and determines the behavior of the cosmic fluid, in other words, the cosmic fluid now can behave as phantom dark energy or as quintessence one. 
\end{abstract}


\maketitle

\section{Introduction}
Several of the physical properties exhibited by macroscopic systems in nature are well described by the standard thermodynamics. However, when we look at objects as black holes and the expanding Universe, the restatement to a thermodynamics description of such phenomena is not self-evident. For our good fortune, the connection between gravity and thermodynamics was given by Bekenstein \cite{bekenstein} based on the Hawking's area theorem \cite{haw}. Since the black hole area, namely $A$, can not decrease therefore Bekenstein conjectured, $S_{BH} \propto A$, being $S_{BH}$ the entropy of the black hole. From the latter proportionality Bekenstein proposed a generalized form for the second law of thermodynamics: {\it The sum of the black hole entropy, $S_{BH}$, and the ordinary entropy of matter and radiation fields in black hole exterior region, $S$, can not decrease}, i.e., $\Delta (S_{BH}+S)\geq 0$. For consistency, this analogy between black holes and non-relativistic systems required black holes to emit thermal radiation like a body with certain absolute temperature \cite{haw2}, nowadays we call to this physical phenomenon Hawking's radiation and is understood in terms of spontaneous creation of pair of particles near the event horizon, such interpretation does not contradict any physical principle. This is a clear example of the scope of thermodynamics. The history of the analogy between black holes and macroscopic systems does not end here, since the entropy is a measure of the available phase space for the system, then an upper bound on the entropy might exist, the bound results to be positive and depends on the size of the system, this was found by Bekenstein in Ref. \cite{bound} and offered the possibility to consider the Universe as a thermodynamical system. In Ref. \cite{horizon} was found that if the particle horizon\footnote{A more recent result shows that for an expanding Universe the apparent horizon differs only by a small quantity from the event horizon. Therefore the thermodynamics description does not change substantially for each case \cite{bousso}.} is used to characterize the radius of the Universe, then the Friedmann model is useful to describe the behavior of the Universe at late times, but, thermodynamics inconsistencies appear at the initial singularity; Bekenstein concluded that such singularity is forbidden by thermodynamics since the upper bound for the entropy is not guaranteed due to the high curvature.\\

It is worth mentioning that a general proof for the second law of thermodynamics (and some possible extensions) conjectured by Bekenstein does not exist so far, for instance, a proof based on fundamental aspects of a quantum gravity theory. However, to our knowledge, there are not physical situations that contradict the validity of this principle. For instance, if we focus on the cosmology scenario, it was found that second law is obeyed by the Universe at large scales \cite{pavon1} and the entropy of the Universe tends to a maximum value \cite{pavon2}, this latter result backs the Bekenstein bound. Other result shows from a quantum perspective that entropy must grow as Universe evolves \cite{omar}, just to mention some. A vast number of works in this topic can be found in the literature for several cosmological models.

Motivated by the validity of the generalized second law and the consistency that represents for a cosmological model from the thermodynamics point of view, in this work we explore the cosmological implications arising from the implementation of the second law for two entropies: the entropy of the apparent horizon and that associated to matter creation effects, i.e., when gravitational particle production is allowed. As far as we know this is the first work where the generalized second law is directly incorporated into the matter creation model. An interesting aspect of matter creation models is their relationship with the so called arrow of time (a distinction of the past from the future). Due to the asymmetry that they introduce in the evolution of the Universe, it is possible to have an agreement between the thermodynamical arrow of time (the one in which entropy increases) and the cosmological arrow of time (direction in which the Universe expands and evolves in time). Therefore, the Universe in this kind of scheme expands in the same direction of its thermodynamics evolution, leading to a consistent thermodynamics description, we refer the reader to Refs. \cite{harko1, harko2}, where these kind of aspects of matter creation models are explored within some specific scenarios for the early and late times Universe.   

Surprisingly we find that in order to maintain the compatibility with thermodynamics, our model admits a phantom-quintessence behavior backed by a chemical potential (which is not common in the literature since in the matter creation approach the crossing to the phantom regime is not obtained frequently, as can be seen in \cite{us2}). Otherwise the Nernst theorem is violated for null chemical potential. We meet with a promising scenario if we consider that phantom dark energy is a viable candidate to solve inconsistencies of the standard cosmological model as the already known $H_{0}$ tension, see for instance Refs. \cite{tension} and \cite{valentino}.

This paper is organized as follows: In Section \ref{sec:creation} we provide a short description for the cosmological scenario of production of particles. We emphasize the meaning and repercussions of the adiabatic condition and its interpretation from previous works. In Section \ref{sec:general} we explore the cosmological consequences from the generalization of the entropy flux vector when the apparent horizon and matter creations effects are included into the second law of thermodynamics. As we will see, a phantom regime is allowed but contradicts the thermodynamics grounds. However, in Section \ref{sec:chemical} we show that the inclusion of chemical potential solves the thermodynamics issues of the model and allows a more general description for the cosmic fluid: a phantom-quintessence behavior is permitted. In Section \ref{sec:final} we give the final comments of our work.    

\section{Production of particles}
\label{sec:creation}

From now on, we will describe our cosmological model assuming a Friedmann-Lemaitre-Robertson-Walker (FLRW) metric with flat space section ($k=0$); units are chosen so that $8\pi G=c=k_{B}=1$, we also consider the standard relation between redshift, $z$, and the scale factor $a$ given as $1+z=a^{-1}$. 

Additionally to Friedman equations, creation effects in the cosmic fluid are characterized by the introduction of a positive term (production of particles), $\Gamma$, into the particle number balance \cite{classic}
\begin{equation}
    n^{a}_{;a} = n\Gamma,
\end{equation}
where $n^{a}:=nu^{a}$, being $u^{a}$ the four-acceleration; for the case of a FLRW geometry the previous equation takes its well-known form
\begin{equation}
    \dot{n}+3Hn = n\Gamma,
    \label{eq:number}
\end{equation}
with Hubble parameter $H$ and $n$ represents the particle number density. The appearance of particles in this context is due to the existence of fluctuations in the geometrical background associated to quantum fluctuations of matter fields, in other words, the energy of produced particles is obtained from the gravitational field, then the following condition is given for the source term, $\Gamma>0$; see Ref. \cite{creation} for a detailed analysis of this formalism. The dot denotes derivatives with respect to cosmic time. The dynamical pressure must be decomposed as
\begin{equation}
    P=p+p_{c},\label{eq:ppc}
\end{equation}
where $p$ is the equilibrium pressure and $p_{c}$ acts as a correction term which appears only within the matter creation scenario and is interpreted as a non-thermal pressure, below we will comment more about this term. 

The consideration of Eq. (\ref{eq:ppc}) leads to the following set of dynamical equations for the fluid 
\begin{align}
& 3H^2 = \rho, \label{eq:fried1}\\
& 2\dot{H} + 3H^2 = -(p+p_c) \label{eq:accel},
\end{align}
where $\rho$ will characterize the energy density of the created matter. The continuity equation for $\rho$ is also modified by the correction term introduced into the total pressure of the fluid
\begin{equation}
    \dot{\rho} + 3H (\rho + p+ p_{c}) = 0. \label{eq:mat}
\end{equation}
On the other hand, the entropy flux vector is defined as 
\begin{equation}
    s^{a} = snu^{a}, \label{eq:flux}
\end{equation}
therefore the second law of thermodynamics can be written as follows
\begin{equation}
    s^{a}_{;a} \geq 0. 
    \label{eq:produc}
\end{equation}
Using the fact that $u^{a}u_{a} =-1$ together with the expansion of the fluid, $u^{a}_{;a} :=\Theta$, given as $3H$ for a FLRW geometry, we can write 
\begin{equation}
    s^{a}_{;a} = n\dot{s} + sn\Gamma, \label{eq:second}
\end{equation}
where Eq. (\ref{eq:number}) was considered. As can be seen, the fulfillment of Eq. (\ref{eq:produc}) is guaranteed only with a positive source term $\Gamma$.\\ 

The Gibbs equation for the entropy per particle is given as
\begin{equation}
    Tds = d\left(\frac{\rho}{n}\right) + pd\left(\frac{1}{n}\right),\label{eq:gibbs}
\end{equation}
from the above equation we can write
\begin{equation}
    nT\dot{s} = \frac{\partial \rho}{\partial T}\dot{T} - \frac{T}{n}\frac{\partial p}{\partial T}\dot{n},\label{eq:per}
\end{equation}
where we have considered the pair $(n,T)$ as independent variables. With the use of Eqs. (\ref{eq:number}) and (\ref{eq:mat}), the evolution equation for the temperature takes the form
\begin{equation}
    \frac{\dot{T}}{T} = -\left(\frac{\partial T}{\partial \rho}\right)\left[3H\left(\frac{p_{c}}{T}+\frac{\partial p}{\partial T} \right)+\frac{\Gamma}{T}\left(\rho+p -T\frac{\partial p}{\partial T}\right) \right],
\end{equation}
inserting the above equation in (\ref{eq:per}) one gets
\begin{equation}
    nT\dot{s} = -(\rho+p)\left[\Gamma +3H\frac{p_{c}}{\rho+p}\right].\label{eq:per2}
\end{equation}
From this latter result is easily found that the values
\begin{equation}
    \Gamma = -3H\frac{p_{c}}{\rho+p} \ \ \ \mbox{or} \ \ \  p_{c} = -\frac{(\rho+p)}{3H}\Gamma, \label{eq:perfect}
\end{equation}
leads to the adiabatic condition $\dot{s} = 0$. Note that the non-thermal pressure labeled as $p_{c}$ is negative, this is the source that produces the accelerated expansion of the Universe when matter creation effects are taken into account. The form given for $p_{c}$ in Eq. (\ref{eq:perfect}) corresponds to an adiabatic process of particle production from quantum vacuum. However, the second law given in Eq. (\ref{eq:second}) has two contributing terms. Despite the entropy per particle is constant, the entropy of the fluid changes due to the increasing number of produced particles, $s^{a}_{;a} = sn\Gamma$. As discussed in detail in several works of Ref. \cite{classic2}, the condition $\dot{s} = 0$ means that immediately after created, the particles are allowed to access a perfect fluid description. In other words, once we adopt the form given in Eq. (\ref{eq:perfect}) for $p_{c}$, we are dealing with a perfect fluid in which the entropy production is possible due to the particle creation.\\ 

Notice that using Eq. (\ref{eq:perfect}) together with (\ref{eq:number}) in (\ref{eq:mat}) allow us to write
\begin{equation}
    \dot{\rho} =\frac{\dot{n}}{n}(\rho+p),
\end{equation}
then if we assume $\rho = mn$ with $m$ being a constant interpreted as the mass of the particle, we get the condition $p=0$ from the above equation, which is consistent for created matter with dark matter equation of state. 

\section{Generalizing the entropy flux vector}
\label{sec:general}

Let us consider that the entropy flux vector (\ref{eq:flux}) can be decomposed in two contributions, i.e.,
\begin{equation}
    \bar{s}^{a} = S_{h}u^{a}+snu^{a}, \label{eq:flux2}
\end{equation}
where $s$ characterizes the entropy emerging from particle creation effects and $S_{h}$ the apparent horizon contribution to the entropy, then if we repeat the procedure described in the previous section the second law will be given for this case as 
\begin{equation}
     \bar{s}^{a}_{;a} = \dot{S}_{h}+3HS_{h}+ns\Gamma + n\dot{s} \geq 0,\label{eq:cons}
\end{equation}
then the election of Eq. (\ref{eq:perfect}) for the correction term $p_{c}$ leads to $\dot{s} = 0$. According to Eq. (\ref{eq:cons}), we are considering a generalized form for the second law as proposed by Bekenstein: the total entropy is given by the contribution of the fluid components of the Universe and that of the horizon.\\ 

For $k=0$, the radius of the apparent horizon in a FLRW spacetime is simply the Hubble horizon
\begin{equation}
    r_{h}=\frac{1}{H},
\end{equation}
and its entropy
\begin{equation}
    S_{h}=\frac{A}{4}, \label{eq:app}
\end{equation}
where $A$ is the area of the apparent horizon, $A=4\pi r^{2}_{h}$. The time derivative for the entropy (\ref{eq:app}) is given directly as $\dot{S}_{h} = 2\pi r_{h}\dot{r}_{h}$ and 
\begin{equation}
    \dot{r}_{h} = -\frac{\dot{H}}{H^{2}} = 1+q=\frac{r^{2}_{h}}{2}(\rho + p + p_{c}), \label{eq:qq}
\end{equation}
being $q:=-1-\dot{H}/H^{2}$, the deceleration parameter as usual. Regard that $\dot{r}_{h}$ can be written also in terms of the energy density and pressure of the fluid by means of the dynamical equations (\ref{eq:fried1}) and (\ref{eq:accel}). In order to visualize the cosmological implications of the creation pressure contribution, we will consider created particles being dark matter satisfying a pressureless equation of state, therefore from now on we will assume $p=0$. From our previous results we can obtain
\begin{equation}
    \dot{S}_{h} = \frac{A}{4}(\rho + p_{c})r_{h} = 3\pi r_{h}\left( 1-\frac{\Gamma}{3H}\right), \label{eq:timehor}
\end{equation}
where we have considered (\ref{eq:perfect}). It is worthy to mention that from the previous expression for $\dot{S}_{h}$ and Eq. (\ref{eq:qq}), we can relate the source term $\Gamma$ and $q$ as follows
\begin{equation}
    q = \frac{1}{2}\left(1-\frac{\Gamma}{H}\right).\label{eq:gammaq}
\end{equation}
This latter result indicates that from the time derivative of the entropy defined on the horizon, we can obtain information about the particle production rate and relate it to some well-known cosmological quantities of the observable Universe, i.e., $q$ is given in terms of $\Gamma$. The deceleration parameter has been estimated in the literature with the use cosmological data for several dark energy models. This relation is relevant since the value at present time of the parameter $q$ fixes the value for $\Gamma$. It is worthy to mention that only for $\Gamma > 0$ the deceleration parameter (\ref{eq:gammaq}) can take negative values, in this sense, the created particles will exhibit a dark energy behavior. Notice that for $\Gamma=0$ we obtain $q=1/2$, this is consistent with the dark matter fluid since we are considering the case $p=0$, therefore the dark energy behavior of this scenario is possible due to the existence of the positive source term $\Gamma$ which implies a negative pressure contribution, $p_{c}<0$, as can be seen from Eq. (\ref{eq:perfect}). 

Besides, if we consider $s^{a}_{;a} = 0$ and the condition $\dot{s} = 0$ we have that $s$ is constant, for thermodynamics consistency such constant must be positive. 

Therefore, by considering the aforementioned conditions and our previous results it is possible to obtain the following expression for the entropy
\begin{equation}
    ns\Gamma = -\dot{S}_{h}-3HS_{h} = 3\pi r_{h}\left(\frac{\Gamma}{3H}-1\right)-3H\pi r^{2}_{h}, \label{eq:produced}
\end{equation}
which takes the form
\begin{equation}
    s = \frac{3\pi}{nH\Gamma}\left(\frac{\Gamma}{3H} - 2\right). \label{eq:entro}
\end{equation}  
From the above equation we can write for the source term the convenient following expression
\begin{equation}
    \frac{\Gamma}{3H} = 2\left[1-\frac{3s}{m\pi}H^{4}\right]^{-1}, \label{eq:imposed}
\end{equation}
where we have considered $\rho = mn$ and the Friedmann constraint (\ref{eq:fried1}), we observe that the condition $\Gamma/3H > 1$ (particle production) can be read directly from the last expression. From Eqs. (\ref{eq:gammaq}), (\ref{eq:entro}) and (\ref{eq:imposed}) the expression for $s$ can be written in terms of $q$, yielding
\begin{equation}
    s = -\frac{\pi}{n H\Gamma}(5+2q) = -\frac{m\pi}{3H^{4}}\frac{(5+2q)}{(1-2q)}. 
    \label{eq:last1}
\end{equation}
As can be seen from Eq. (\ref{eq:last1}), in order to have a positive constant for the entropy at present time, we are restricted to the condition, $q < -5/2$, which is consistent with an accelerated Universe; the value of the deceleration parameter only admits an over accelerated stage for the Universe, i.e., phantom regime $(q<-1)$.\\

As can be seen in Eq. (\ref{eq:produced}), the entropy of the produced matter can be related to the apparent horizon entropy. However, for an accelerated cosmic expansion there exists an outward flux of matter fields through the horizon due to the difference between the comoving expansion speed and that of the formation of apparent horizon, as established in Ref. \cite{flux}. Then, for a more robust thermodynamics description, the total entropy production given in (\ref{eq:produced}) must take into account the rate of change of the particle number inside the horizon which is given by a term of the form \cite{flux}
\begin{equation}
    \dot{N} \propto \left(\frac{4\pi}{H^{2}} \right)q = \left(\frac{2\pi}{H^{2}} \right)\left(1-\frac{\Gamma}{H} \right), \label{eq:flux} 
\end{equation}
where we have considered Eq. (\ref{eq:gammaq}) and $N$ is the total number of particles contained in the volume $V$, i.e., $n=N/V$; then $\dot{N}$ becomes negative due to the accelerated expansion since $q<0$, or equivalently, outward flux through the horizon. In case of consideration of the aforementioned flux, the Eqs. (\ref{eq:entro}) and (\ref{eq:last1}) will have an extra contribution which in consequence modifies the bound for the deceleration parameter obtained from Eq. (\ref{eq:last1}). As probed in \cite{flux} for different horizon entropies during the early stage of the universe, the matter flux given by (\ref{eq:flux}) validates the second law of thermodynamics. For our purposes we focus solely on the cosmological implications of the phantom scenario emerging from the consideration of matter production inside the horizon since this is not common in this kind of cosmological model.\\ 

In the following we show that the Hubble parameter of the model allows a future singularity, this behavior is in agreement with a Big Rip cosmology and is consistent with the value obtained above for $q$. Using the expression given in (\ref{eq:last1}) and the definition $\eta:= m\pi/(3s)$, we can write the deceleration parameter as follows
\begin{equation}
    q\left( H\right) =\frac{1}{2}\left(1+\frac{5\eta}{H^{4}}\right)\left(1-\frac{\eta}{H^{4}}\right)^{-1},
\end{equation}
which in turn results in a differential equation for the Hubble parameter if we use the standard definition for $q$ given previously in Eq. (\ref{eq:qq})
\begin{equation}
    \frac{d}{dt}\left( \frac{\eta ^{1/4}}{H}\right) = \eta ^{1/4}\left[ 1+\frac{1}{2}\left( 1+5\frac{\eta }{H^{4}}\right) \left( 1-\frac{\eta }{H^{4}}\right) ^{-1}\right],
\end{equation}
whose solution reads
\begin{equation}
    H(t) = \frac{H_{0}}{1-\frac{3}{2}H_{0}(t-t_{0})} = H_{0}(t-t_{s})^{-1},\label{eq:singu}
\end{equation}
in the limit $\eta^{1/4}/H \ll 1$ and we have defined the Hubble constant, $H(t=t_{0})=H_{0}$, at initial time $t_{0}$. 

The Hubble parameter diverges at a specific time at the future, $t=t_{s}$, given as $t_{s} = t_{0} + 2/(3H_{0})$, i.e., this is consistent with the value $q<-5/2$ obtained from the positivity condition for the entropy, $s$. According to Eq. (\ref{eq:imposed}) one gets
\begin{equation}
    \frac{\Gamma}{3H} = 2\left[1-\frac{H^{4}}{\eta}\right]^{-1} < 0,
\end{equation}
since $\eta > 0$ and $H$ diverges as $t\rightarrow t_{s}$, as shown above. Then this phantom scenario is not backed by thermodynamics because leads to particle annihilation scenario and negative entropy production, this latter condition can be seen from Eq. (\ref{eq:second}).\\ 

Below we discuss an extension for the formalism developed in this section in order to solve some of the controversies discussed above.  

\section{Solving thermodynamics issues with chemical potential}
\label{sec:chemical}

An interesting feature of the scheme discussed previously is the fact of the emerging phantom scenario from the simple consideration of created particles being trapped by the apparent horizon, this kind of bulk-boundary {\it interaction} mediated by a single fluid has been subject of investigation and deserves more attention, see for instance \cite{us}. As can be found in the literature, see references given in \cite{chemical}, the chemical potential, usually termed as $\mu$, has been useful to solve some inconsistencies appearing in the thermodynamics description of phantom dark energy. See also Ref. \cite{chemical2}, where the negativity of chemical potential plays a relevant role for a consistent thermodynamics description of a phantom scenario.\\ 

It is worthy to mention that from the thermodynamics point of view we are not introducing new degrees of freedom with the inclusion of chemical potential in our description since we are dealing with a single fluid, therefore the chemical potential is simply the Gibbs free energy per particle see Refs. \cite{callen, reif} for a detailed discussion; when chemical potential is included the Gibbs equation takes the following form
\begin{equation}
    Tds = d\left(\frac{\rho}{n}\right) + pd\left(\frac{1}{n}\right)-\mu d\left(\frac{1}{n}\right).\label{eq:chem1}
\end{equation}
If we repeat the procedure described in section \ref{sec:creation}, we can write the following expression for the entropy production
\begin{equation}
    nT\dot{s} = -(\rho+p-\mu)\left[\Gamma +3H\frac{p_{c}+\mu}{\rho+p-\mu}\right].\label{eq:chem2}
\end{equation}
Notice that in this case the adiabatic condition leads to the condition
\begin{equation}
    \frac{\Gamma}{3H} = -\frac{(p_{c}+\mu)}{\rho + p - \mu} \ \ \ \mbox{or} \ \ \  p_{c} = - \mu - \frac{(\rho+p-\mu)}{3H}\Gamma,
\end{equation}
thus in this scenario the creation pressure can not be related directly to the source term as in the previous analysis since now we are including the chemical potential, as can be seen, the chemical potential also contributes to cosmic expansion. 

In this case the acceleration equation (\ref{eq:accel}) it is useful to provide an expression for $p_{c}$ in terms of some relevant cosmological parameters, yielding
\begin{equation}
    p_{c}= -H^{2}(1-2q),
\end{equation}
as expected, such pressure appears to be negative for accelerated expansion, we have considered $p=0$. Similarly to the case studied previously, we can write the following expression for the entropy of the created matter sector
\begin{equation}
    s = -\frac{m\pi}{3H^{2}}\left\lbrace \frac{(5+2q)\left(1-\frac{\mu}{3H^{2}}\right)}{H^{2}(1-2q)-\mu}\right\rbrace, \label{eq:chem3} 
\end{equation}
which represents the generalization of Eq. (\ref{eq:last1}) when $\mu \neq 0$. It is worthy to mention that in this case due to the presence of chemical potential the positivity of the entropy can be maintained relaxing the condition on the value of the deceleration parameter, i.e., for an accelerated Universe the conditions $\mu > 3H^{2}$ and $q > -5/2$ are sufficient, for this cosmological scheme the phantom regime is allowed as well as quintessence. However, the value $q=-1$ (which represents a dark energy modeled by a cosmological constant) is excluded at present time, note that in such case the entropy is negative for $\mu > 3H^{2}$.

The exclusion of the value $q=-1$ at present time is consistent if we appeal to the conditions that must be satisfied by the area of the apparent horizon in order to reach the thermal equilibrium at the far future ($z\rightarrow-1$ or $t\rightarrow \infty$), such conditions are $A'\geq 0$ at any time and $A'' \leq 0$ at late times, where the prime denotes derivatives with respect to the scale factor. If we calculate the derivatives of the area given in the entropy Eq. \ref{eq:app} we obtain, $A' = 2A[(1+q)/a]$ and $A''=2A[2((1+q)/a)^{2}-(1+q)/a^{2}+q'/a]$, then the condition $q\rightarrow-1$ for an evolving deceleration parameter is possible only at late times together with $q'<0$, these conditions for the area of the apparent horizon were discussed in detail in Ref. \cite{second}.\\ 

The generalized form of Eq. (\ref{eq:imposed}) with $\mu \neq 0$ is given by
\begin{equation}
     \frac{\Gamma}{3H} = \left(2-\frac{\mu}{3H^{2}}\right)\left[1-\frac{3s}{m\pi}H^{4}-\frac{\mu}{3H^{2}}\right]^{-1}. \label{eq:imposed2}
\end{equation}
As in the previous section, the source term is also related to the deceleration parameter (see Eq. (\ref{eq:gammaq})), in this case we have
\begin{equation}
    \frac{\Gamma}{3H} = \frac{(1-2q)-\frac{\mu}{H^{2}}}{3\left(1-\frac{\mu}{3H^{2}}\right)}.\label{eq:chemi}
\end{equation}
Using these latter expressions for the source term we can write an expression that involves the Hubble parameter
\begin{equation}
    \frac{1}{3}(1-2q) =1+\frac{2}{3}\frac{\dot{H}}{H^{2}} = x+(x-1)(x-2)\left[1-x-\frac{\beta^{2}}{x^{2}}\right]^{-1}, \label{eq:chem4}
\end{equation}
where we have defined $x:=\mu/3H^{2}$ and $\beta := \mu/3\sqrt{\eta}$ with $\eta$ defined as in the previous section. However, the differential equation given above is difficult to solve analytically. Next we specialize it to the phantom scenario in order to illustrate the thermodynamics of the model.\\

$\bullet$ {\bf Phantom case}\\

According to the condition obtained for $q$ from Eq. (\ref{eq:chem3}), the phantom regime is admissible, then we can consider the following Ansatz for the Hubble parameter if we focus on the aforementioned scenario \cite{ansatz}
\begin{equation}
    H(t)= \sigma (t_{s}-t)^{-1},\label{eq:ansatz}
\end{equation}
where $\sigma$ is a positive constant in order to have an expanding Universe and $t_{s}$ represents a time at the future at which the singularity takes place. Notice that this Ansatz resembles the analytical solution (\ref{eq:singu}) obtained in the scenario discussed previously. If we insert the Ansatz for the Hubble parameter in Eq. (\ref{eq:chem4}) we can obtain the following expression for the chemical potential
\begin{equation}
    \mu(t) = 3\sigma^{2}\left(1+\frac{2\sigma^{4}}{3\sigma^{5}-(2-3\sigma) (t_{s}-t)^{4}\eta} \right)(t_{s}-t)^{-2}, \label{eq:potential} 
\end{equation}
which diverges at $t=t_{s}$ meaning that the Gibbs free energy per particle (energy density of the fluid) diverges, this is a typical behavior of a Big Rip singularity, besides the chemical potential remains positive. Therefore from (\ref{eq:chemi}) one gets 
\begin{equation}
    \frac{\Gamma}{3H} = \frac{(2-3\sigma)(t_{s}-t)^{4}\eta}{3\sigma^{5}},\label{eq:chemi2}
\end{equation}
then the condition $0 < \sigma < 2/3$ guarantees the positivity of the source term and leads to an expanding Universe. As can be seen from the previous expression, the matter creation ceases in the limit, $t\rightarrow t_{s}$. From eqs. (\ref{eq:ansatz}) and (\ref{eq:potential}) we can compute the following quotient
\begin{equation}
    \frac{\mu(t)}{3H^{2}(t)} = 1+\frac{2\sigma^{4}/3}{\sigma^{5}-(2/3-\sigma)(t_{s}-t)^{4}\eta} > 1, \label{eq:quotient} 
\end{equation}
as required by the entropy (\ref{eq:chem3}) to be positive and takes the positive value $1+2/(3\sigma)$ at $t=t_{s}$. Then the inclusion of chemical potential in our description leads to a well-defined phantom regime from the thermodynamics point of view.

\section{Final Remarks}
\label{sec:final}

In this work we concentrated on the generalized form of the second law of thermodynamics and the cosmological implications when the apparent horizon contribution is included within the particle production picture, as far as we know this is the first work where this situation is considered. 

An interesting feature of this cosmological model is that the entropy production is due entirely to the increasing number of particles in the fluid. A first result of this setup is that in order to maintain the positivity of the entropy (Nernst theorem) associated to the produced particles, the cosmic fluid must behave as phantom dark energy; the value of the deceleration parameter is restricted to the region $q < -5/2$. However, under these considerations the reconstructed source term, $\Gamma$, is negative. This leads to thermodynamics inconsistencies as negative entropy production, in this regard; the modification of physical quantities associated to the apparent horizon seems to be a first approach to solve such inconsistencies. For instance, if we adopt the Barrow proposal for the entropy \cite{barrow}, the deceleration parameter is now restricted to the region
\begin{equation}
    q < - \left(\frac{5+\Delta}{2+\Delta} \right),
\end{equation}
where $\Delta$ is the Barrow exponent with values $0 \leq \Delta \leq 1$. Notice that in this case as $\Delta \rightarrow 1$ we obtain $q < -2$, implying that the evolution of the Universe is strongly influenced by the horizon description but the thermodynamics inconsistencies in the model can not be removed using this avenue. Besides, we must take into account that not all generalizations for entropy respect the second law of thermodynamics, see for instance Ref. \cite{saribarrow}, where it was found that in the presence of the Barrow entropy the generalized second law is violated. However, the consideration of entropy generalizations as the proposals given in Ref. \cite{generalized} could lead to a viable description of our model since such generalizations provide a consistent unified description of early inflation and late dark energy with few free parameters. We leave this subject open for future investigation.

In order to solve the thermodynamics disagreements of the model, we opted to include chemical potential in our description. After the inclusion of chemical potential, as first result we observed that the positivity condition for the entropy of created matter relies strongly on the value of $\mu$, such condition is $\mu > 3H^{2}$; therefore the condition for the accessible values of the deceleration parameter can be relaxed, in this case we obtain $q>-5/2$; therefore this condition allows a phantom-quintessence behavior for the cosmic fluid, see for instance Ref. \cite{uscm} where even with inclusion of holographic dark energy, the phantom regime is not accessible. Due to the complexity of the resulting equations in the formalism it is no longer possible to obtain analytical solutions for the Hubble parameter as in the case of null chemical potential. However, under the consideration of a well-known Ansatz of the Hubble parameter for the phantom regime, we reconstructed explicit functions of cosmic time for $\mu$ and $\Gamma$; while the chemical potential is positive and diverges as we approach to the future singularity, the source term tends to zero, i.e., the matter creation decreases near the future singularity. According to the conditions obtained in this construction, the $\Gamma$ term is positive and does not contradict the principles of thermodynamics. Besides, the condition $\mu > 3H^{2}$ is satisfied and in consequence the entropy of the created dark energy is also positive. We would like to remark the fact that chemical potential is not an extra degree of freedom since we are dealing with a single fluid, in such case the chemical potential is simply the Gibbs free energy per particle.

A remarkable virtue of this framework is the gain of the reconstructed source term $\Gamma$ in each case explored (Eqs. (\ref{eq:imposed}) and (\ref{eq:imposed2})), this is not common in the literature. See for instance Ref. \cite{us2}, where the phantom-quintessence behavior is accessible only under the consideration of {\it ad hoc} forms for $\Gamma$. We also highlight the fact that in this cosmological model the condition of thermal equilibrium at late times $(z\rightarrow -1)$ can not be fulfilled for the phantom regime since the model diverges at some time at the future $(t=t_{s} \ \ \mbox{which corresponds to} \ \ z>-1)$, nevertheless this is consistent with some other results, see for instance \cite{equili}, where it was found that the achievement of thermal equilibrium between dark energy and the horizon depends on very restrictive conditions: the nature of the dark energy and the content trapped by the horizon. 
Finally, we emphasize that there are still a lot of subtle issues worth investigating when tests against observations are implemented for this kind of scenario. However, we feel that any further exploration along this line would justify a separate analysis.

\section*{Acknowledgments}
MC work has been supported by S.N.I.I. (CONAHCyT-M\'exico). VHC work was partially supported by the Center Cefitev UV. MC appreciates clarifying comments in thermodynamics made by C. Contreras.\\

$\bullet$ {\bf Data Availability Statement}\\ 
No datasets were generated or analysed during the current study.

\end{document}